\documentclass[aps,twocolumn,floatfix,tightenlines,prb,superscriptaddress]{revtex4-2}
\usepackage{graphics}
\usepackage{subcaption}
\usepackage{epstopdf}
\usepackage{epsfig}
\usepackage{graphicx}
\usepackage{epsf,epic}
\usepackage{color}
\usepackage{amsmath}
\usepackage{booktabs}
\usepackage{multirow}
\usepackage{physics}
\usepackage{hyperref}
\usepackage{amsfonts}
\usepackage{wrapfig}
\usepackage{breqn}
\usepackage{pstricks}
\usepackage[utf8]{inputenc}
\usepackage{multirow}
\usepackage{fancyref}
\usepackage{pst-node}
\usepackage{soul}
\usepackage{float}
\usepackage{bm}
\usepackage{dcolumn}
\newcommand{\etal}{\textit{et al.\ }}

\makeatletter
\usepackage{etoolbox} 
\appto{\appendix}{%
	\@ifstar{\def\theequation@prefix{A.}}%
	{}%
}
\preto\maketitle{%
  \begingroup\lccode`~=`,
  \lowercase{\endgroup
  \let\saved@breqn@active@comma~
  \let~}\active@comma 
}
\appto\maketitle{%
  \begingroup\lccode`~=`,
  \lowercase{\endgroup
  \let~}\saved@breqn@active@comma 
}
\makeatother
\begin{document}
\title{Magnetic exchange interactions in the molecular orbital spin system NaV$_2$O$_5$ from a distributed moment point of view.}
\author{Ilteris K. Turan}
\affiliation{Department of Physics, Case Western Reserve University, 10900 Euclid Avenue, Cleveland, Ohio 44106-7079, USA}
\author{Claudio Garcia}
\affiliation{Department of Physics, Case Western Reserve University, 10900 Euclid Avenue, Cleveland, Ohio 44106-7079, USA}
  \affiliation{National Laboratory of the Rockies, Golden, Colorado 80401, USA}
  \author{Jerome Jackson}
  \affiliation{Scientific Computing Department, STCF Daresbury Laboratory, Warrington WA4 4AD, United Kingdom}
  \author{Swagata Acharya}
  \affiliation{National Laboratory of the Rockies, Golden, Colorado 80401, USA}
  \author{Mark van Schilfgaarde}
  \affiliation{National Laboratory of the Rockies, Golden, Colorado 80401, USA}
 \author{Walter R L. Lambrecht}
\email{walter.lambrecht@case.edu}
\affiliation{Department of Physics, Case Western Reserve University, 10900 Euclid Avenue, Cleveland, Ohio 44106-7079, USA}
  \begin{abstract}
The magnetism  in NaV$_2$O$_5$ results from doping of the narrow split-off conduction band of V$_2$O$_5$, which becomes half-filled and leads to a  Mott-insulating splitting of spin resolved bands with antiferromagnetic order of the spins along the zigzag chains. In the $Pmmn$ structure the spin of this $S=1/2$  system is equally shared between two vanadium atoms, residing in a molecular orbital type state. While below 34 K  a charge disproportionation occurs into V$^{4+}$ and V$^{5+}$, the situation above this temperature is less clear and amounts to a fluctuating moment with equal probability of occupancy of each V. While traditionally described as a quarter-filled ladder system with electron spin localized on the rungs of the ladder, we here take a {\sl distributed moment} approach in terms of the spin density lumped into  individual atomic magnetic sites, including the small induced moments on the oxygen atoms. Exchange interactions are calculated between these sites using a linear response approach based on quasiparticle-self-consistent $GW$ band structures.  Surprisingly we find the exchange interactions between the small magnetic moments induced on the vanadyl and bridge oxygen sites to be  larger than the exchange interactions between vanadium atoms. Their role in the critical temperature is found to be crucial. The spin wave spectra obtained from this  classical Heisenberg type Hamiltonian  contains unusual optic spin wave type collective excitations of high energy, the significance of which is discussed.
\end{abstract}
\maketitle
Since the claim by Isobe and Ueda \cite{Isobe1996} in 1996 that $\alpha^\prime$-NaV$_2$O$_5$ undergoes a spin-Peierls transition at 34 K, it became the subject of intense study.  $\alpha^\prime$-NaV$_2$O$_5$ maintains essentially the unique layered crystal structure of V$_2$O$_5$ with Na intercalated between the layers and located in the center of the large interstitices between the double zigzag chains. With Na donating one electron per V$_2$O$_5$ unit, it was at that time believed that charge disproportionation occurs into V$^{4+}$ and V$^{5+}$ arranged in alternate atomic chains resulting in the $P2_1mn$ space group. The isolated chains with antiferromagnetic superexchange along the chains in the {\bf b}-direction was then assumed to explain the broad maximum in magnetic susceptibility at 320 K, measured earlier by Carpy \etal \cite{Carpy72} and led to an estimate of the antiferromagnetic exchange interaction between nearest neighbors along the chain of about $-24$ meV based on the finite chain simulations of Bonner and Fisher\cite{Bonnerfisher64}. A spin-Peierls transition had been predicted in 1D antiferromagnetic Heisenberg chains by Pytte\cite{Pytte74}. However, in 1998, improved X-ray diffraction studies \cite{VonSchnering98,Smolinski1998} showed that at room temperature, the crystal structure was $Pmmn$ and maintained the mirror planes which guarantees all V to be equivalent and thus nominally having a 4.5 valence. This appeared to invalidate the spin-Peierls claim as the chains were now no longer isolated and also made the existence of a band gap puzzling. This generated a spur of interest, \cite{Horsch1998,Ueda2012,Konstantinovic2001,Mostovoy1999,Cuoco1999,Presura2000,Long1999} which eventually was settled with the realization that the electron donated by Na resides in a molecular type orbital on the rungs of the ladder centered on the bridge oxygen O$_b$ between two equal V. The 34 K transition was then re-interpreted as a charge disproportionation transition but with a more complex zigzag type ordering of V$^{4+}$ and V$^{5+}$ with a $2\times2\times4$ supercell \cite{Fujii97,Ludecke1999,deBoer2000} as evidenced by
NMR spectroscopy\cite{Ohama1999} and anomalous X-ray diffraction \cite{Nakao2000} and supported by theoretical many-body-theory models such as the  $t-J-V$ model \cite{Horsch1998,Cuoco1999} and dynamical mean field theory (DMFT) \cite{Mazurenko2002}. However, much less attention has been devoted to understand the magnetic behavior above the charge disproportionation temperature, in particular whether the broad maximum in susceptibility at 320$\pm50$ K can be interpreted as a N\'eel magnetic ordering temperature and whether its value can be explained by first-principles calculations.

\begin{figure}
(a) \includegraphics[width=6cm]{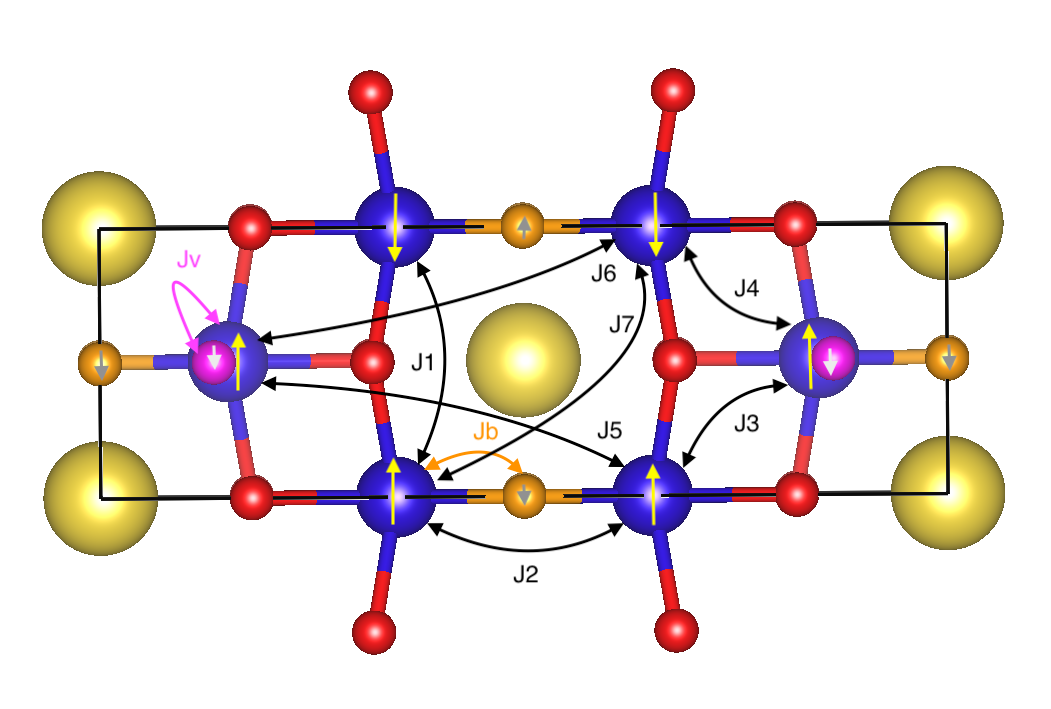}{}
(b)\includegraphics[width=6cm]{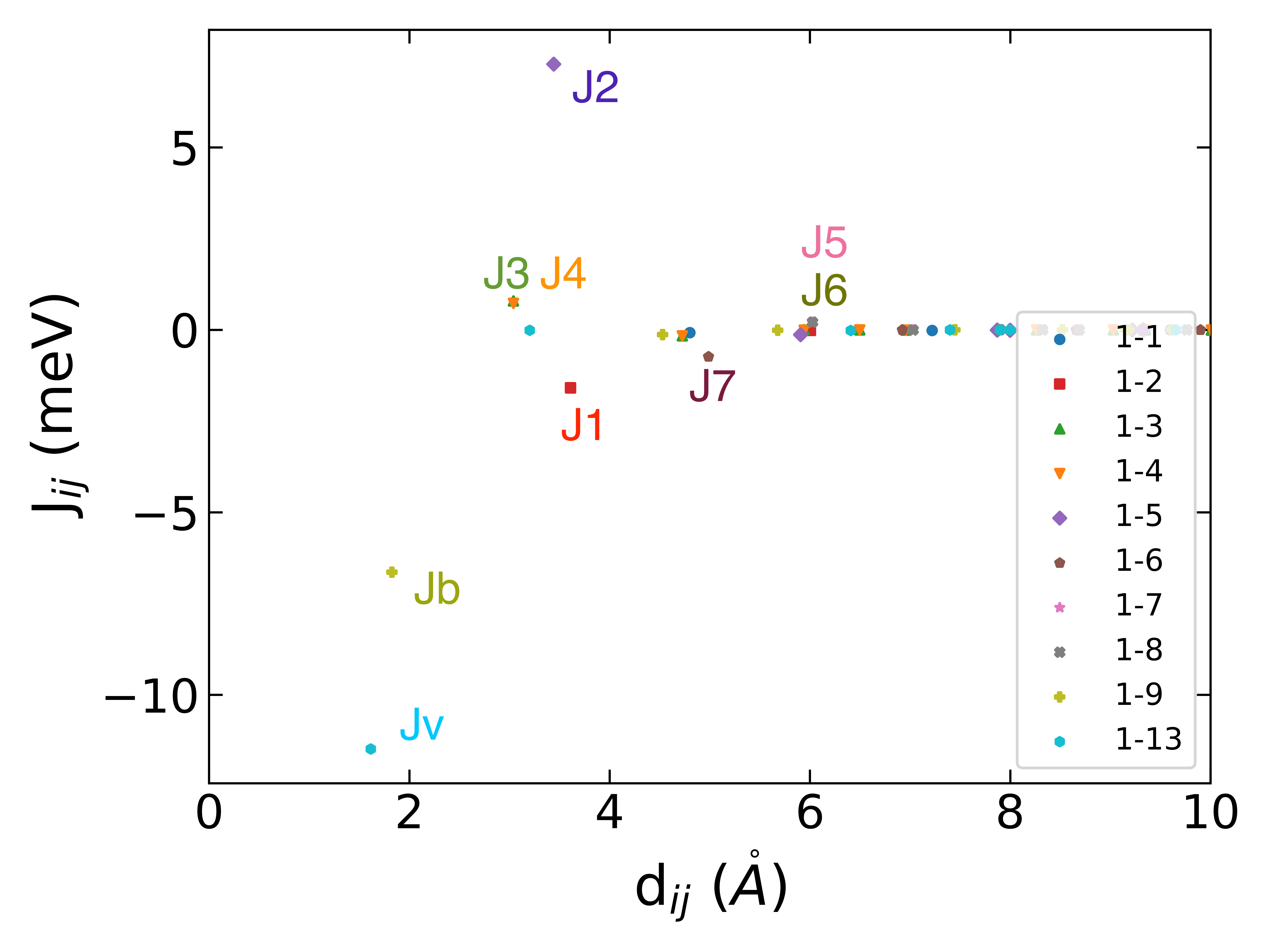}{}
     \caption{
(a) $Pmmn$ crystal structure with spin ordering and labeling of exchange interactions (blue spheres V, yellow Na, red O$_c$, magenta O$_v$, orange O$_b$ (figure made with VESTA3 software \cite{VESTA}, (b) Exchange interactions as function of distance. \label{fig-nav2o5-pmmn}}
\end{figure}

In this paper we focus on the magnetic properties above the charge disproportionation transition and take a {\sl distributed moment} point of view in which magnetic moments are assigned to individual atomic sites and their exchange interactions are calculated from first-principles. To do this, we use the linear response approach of Kotani and van Schilfgaarde\cite{Kotani2008} which has the advantage that it does not make any {\sl a-priori} assumptions on the range of the exchange interactions and is based on accurate quasiparticle band structures obtained in the quasiparticle self-consistent (QS) $GW $ method \cite{MvSQSGWprl,Kotani07} in which $G$ is the one-particle Green's function and $W$ the screened Coulomb interaction as introduced by Hedin\cite{Hedin65,Hedin69}. Like other linear response methods, it is based on the realization that spin wave excitations are the poles of the transverse spin susceptibility $\chi^{+-}({\bf q}, \omega)$, which can be obtained from the band structures, but goes a step further in deriving a site-by-site transverse susceptibility matrix $D^{+-}_{ab}({\bf q},\omega)$ by  integrating over the spheres at atomic sites weighted by a projector function proportional the local magnetization density in the sphere. Inverting this matrix and taking the static approximation $\omega=0$ was shown  in \cite{Kotani2008} to map the problem to a classical Heisenberg type model which directly provides real space exchange interactions by an inverse Fourier transform without the need to fit the spin wave dispersions. The method was shown to provide good agreement for spin wave spectra compared to other approaches and with inelastic neutron scattering experiments when based on QS$GW$ band structures.
A classical Heisenberg model can be  justified for the spin dynamics based on adiabatic decoupling of the longitudinal and transverse spin degrees of freedom \cite{Antropov1996}.
This requires the energy scale of the longitudinal spin response, set by the difference between up and down spin states on the same atom (which here is of order 2 eV), to be larger than the energy scale for transverse response, set by the spin wave energies, as discussed later. It further requires that well defined moments exist in each muffin-tin sphere
with a fixed local spin orientation of the spin density inside the sphere but does not require that this magnetic moment is twice an integer or half-integer spin. 

The spin ordering is shown in Fig.\ref{fig-nav2o5-pmmn}a and is seen to be antiferromagnetic along the {\bf b} direction but ferromagnetic  for the two equivalent V across a bridge. 
The magnetic moment on the V atoms is 0.424 $\mu_B$, 0.065  $\mu_B$ on O$_b$ and 0.042  $\mu_B$ on the O$_v$ and negligible on  O$_c$ (0.00147 $\mu_b$) and Na (0.00028 $\mu_B$). The moments on the O$_v$ and O$_b$  are found  to be oppositely aligned to those on their neighboring V. The most important exchange interactions are labeled.
The exchange interactions, calculated including V, O$_v$, O$_b$ but not O$_c$, Na moments, are summarized in Table \ref{tabexchange-na} column 3
and shown in Fig. \ref{fig-nav2o5-pmmn}b as function of distance.
Surprisingly, we find that the exchange interactions between vanadyl oxygen O$_v$ and bridge oxygen O$_b$ with their nearest neighbor vanadiums are larger than the exchange interaction between vanadium atoms even though the magnetic moments induced on the oxygen are an order of magnitude smaller than the vanadium magnetic moments. Note that  the exchange interactions are here referred to the Heisenberg Hamiltonian written as
$H=-\sum_{ab{\bf T}}J_{a,b+{\bf T}}\bm{e}_a\cdot\bm{e}_{b+{\bf T}}$, with respect to unit vectors and hence absorb
the size of the magnetic moments in the exchange interactions $J$.  Here $a,b$ label sites in the unit cell and ${\bf T}$ labels lattice vectors. These large exchange interactions between O and V reflect the strong antiparallel locking of the moments in these atoms to each other originating in their strong covalent interaction. 

\begin{table}
   \caption{Exchange interactions in NaV$_2$O$_5$ in meV  as defined with respect to the classical Heisenberg Hamiltonian with unit vectors. $N_e$ is the number of equivalent interactions.} \label{tabexchange-na}
   \begin{ruledtabular}
     \begin{tabular}{cccc}
       $J_n$&$N_e$&\multicolumn{1}{c}{$Pmmn$}&\multicolumn{1}{c}{$P2_1mn$} \\ \hline
       $J_1$&2& -1.58& -2.44\\
       $J_1^\prime$&2& & -0.20 \\
       $J_2$&1& 7.35 & 3.88 \\
       $J_3$&1& 0.80 & 0.30 \\
       $J_4$&1& 0.73 & 0.27 \\
       $J_5$&2& 0.21 & 0.21 \\
       $J_6$&2& 0.21  & 0.21 \\
       $J_7$&2& -0.73 & 0.12 \\
       $J_v$&1& -11.53 & -22.39 \\
       $J_b$&1& -6.65 & -2.78\\
     \end{tabular}
   \end{ruledtabular}
\end{table}

The mean field N\'eel temperature can be calculated by constructing the matrix
with elements $J_{ab}({\bf q})=\sum_{\bf T}J_{a,b+{\bf T}}e^{i{\bf q}\cdot{\bf T}}$ at ${\bf q}=0$ and omitting the on-site $J_{a,a}$ term and finding its highest eigenvalue $j_{max}$, which then gives $T_N^{MF}=(2/3k_B)j_{max}$.
This gives a value of 164.8 K. Including only the most important interactions,
one can show that $j_{max}\approx\frac{1}{2}\left[-2J_1+J_2+\sqrt{(-2J_1+J_2)^2+8J_b^2+4J_v^2}\right]$,  which gives $T_N^{MF}=161.7$ K.
On the other hand, when only V exchange interactions are included, this reduces to  $-2J_1+J_2$, which gives a value of only  80.9  K. 

At this point we recall that overall the system  is a $S=1/2$ system for which the quantum correction factor of the critical temperature $(S+1)/S$ is 3, so our estimated mean field temperature for the $Pmmn$  structure is 494 K, which is larger than the experimental $T_{max}$ of 320 K \cite{Carpy72,Isobe1996} by a factor 1.54. Given that mean field theory typically overestimates critical temperatures by a similar factor for 2D systems, this is reasonable.
Using the Random Phase Approximation (RPA) approach \cite{Tyablikov1959,Callen63,Rusz2005} we obtain a $T_N^{RPA}=120$ K and applying the same quantum correction, this gives 360 K. 
The necessity of the quantum correction factor was also pointed out for other small spin systems like NiO in \cite{Kotani2008} and more generally the importance of using a quantum statistical treatment of the finite temperature spin dynamics  is noted in Barker \etal \cite{Barker2019,Barker2021}.

Next, we consider the $P2_1mn$ structure, which exhibits charge disproportionation between V$^{4+}$ and V$^{5+}$. The magnetic moments in this case are 0.683 $\mu_B$ on V$^{4+}$ and 0.174 $\mu_B$ on V$^{5+}$ while they are 0.064 $\mu_B$ on the O$_v$ bound to  V$^{4+}$ and negligble on the one bound to  V$^{5+}$
and 0.048  $\mu_B$ on O$_b$.  In this case, we need to distinguish $J_1^\prime$ between V$^{5+}$ from $J_1$ between V$^{4+}$. The exchange interactions are given in column 4 of Table \ref{tabexchange-na}. We note that in the $P2_1mn$ structure the O are pushed away from the V$^{4+}$ and the Na is attracted and displaced from its position on the mirror plane. This indicates that the charge disproportionation is polaronic in origin and related to the displacement of the atoms around it. The charge disproportionation is not 100 \%. The ratio of the small to large moment is only 0.25, in reasonable agreement with an empirical estimate of 0.38 in \cite{Valenti2001} based on the ratio of the exchange interactions above and below the charge disproportionation temperature.  
The $J_1$ is increased by a factor 1.5 but the $J_2$ is reduced by a factor 0.5. $J_b$ is also reduced because one of its V neighbors is now a V$^{5+}$ but most importantly, the already large $J_v$ is almost doubled. The $T_N^{MF}$  corrected by the same factor 3 now becomes 582 K.  However, in the RPA and including the quantum correction factor,  a much smaller temperature of only 100$\pm10$ K is obtained.  This reflects the almost 1D character of this system, where the MFA gives a larger overestimate. 
While the $P2_1mn$ structure is now believed not to be the low temperature (below 34K) ground state of NaV$_2$O$_5$ ---which has a more complex zig-zag pattern of charge disproportionation and a  spin-Peierls type dimerization along the chains related to the quantum mechanical formation of spin singlets in the dimers--- it still provides a reasonable model to evaluate  the effects of charge disproportionation on the exchange interactions. However is  is unrealistic in yielding isolated 1D chains. Since the $P2_1mn$ structure has lower energy, one may think of the XRD determined $Pmmn$ structure as representing the time average of a randomly fluctuating charge disproportionated structure.

\begin{figure*}[ht]
  \begin{subfigure}{9cm}\centering
    \includegraphics[width=9cm]{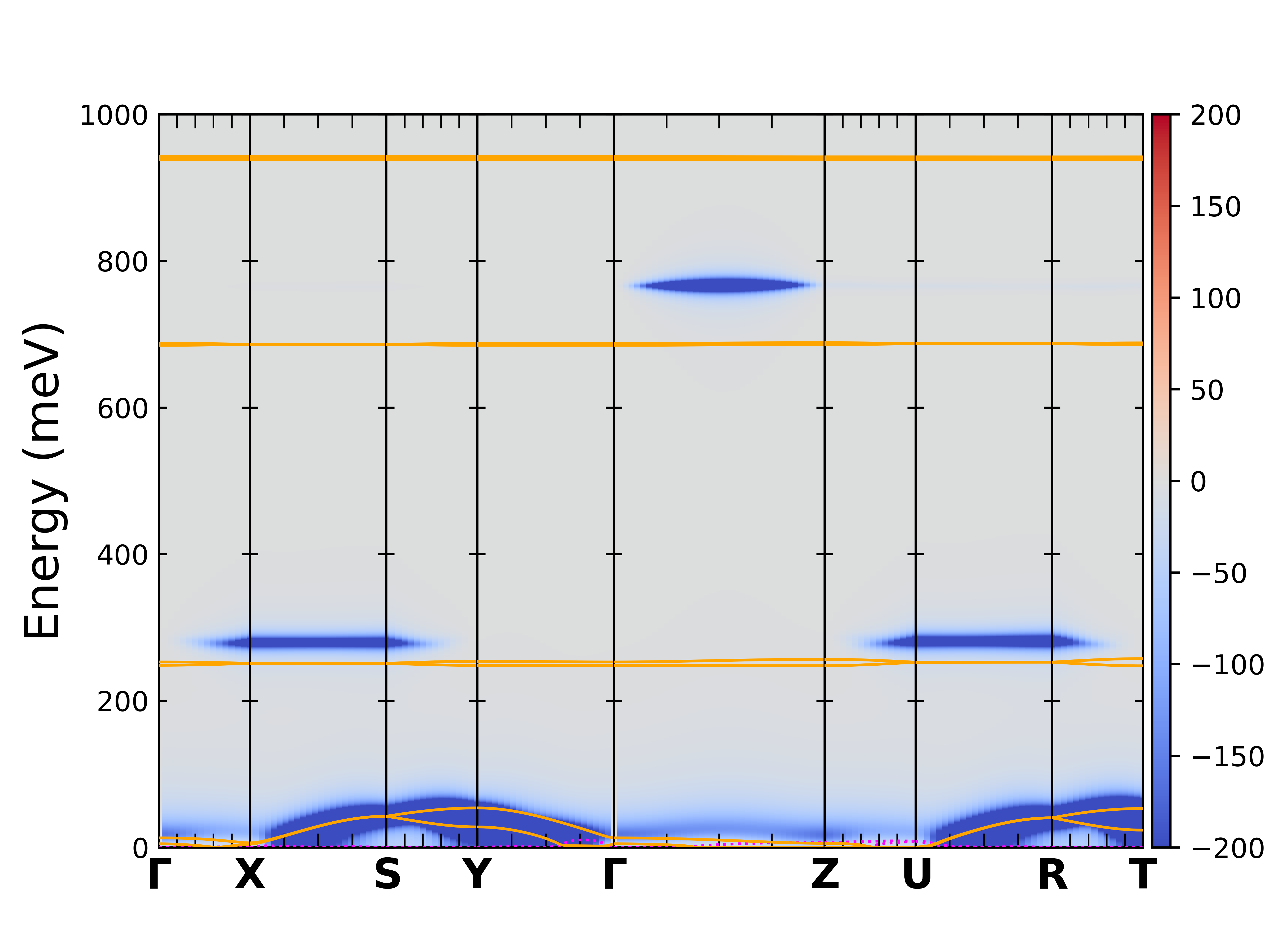}{}
    \caption{}
  \end{subfigure}
  \begin{subfigure}{7cm}
    \includegraphics[width=7cm]{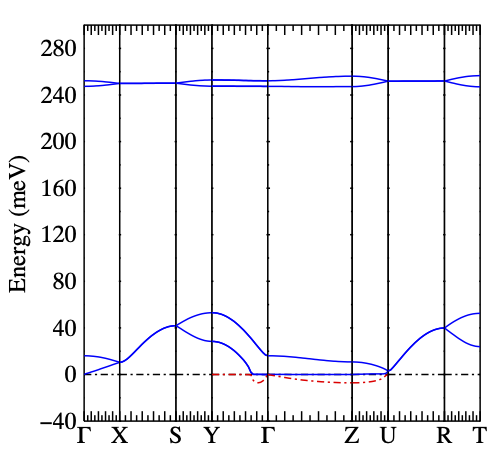}
    \caption{}
  \end{subfigure}
  \caption{(a) Spin waves dispersion (yellow lines) and ${\rm Im}[\chi^{+-}({\bf q},\omega)$ as a heat map of NaV$_2$O$_5$ $Pmmn$; (b) zoom in on low energy range and indicating imaginary eigenvalues in red dashed line} \label{figspinwave}
\end{figure*}

Next, we discuss the spin waves obtained within this Heisenberg model,
shown in Fig. \ref{figspinwave}a.
The heatmap shows peaks at the poles of the dressed $\chi^{+-}$ which are increasingly shifted upward from the spin wave eigenvalues which are obtained from $\chi^{0+-}$. The most striking feature of this spin wave spectrum is the occurrence of several high energy and almost dispersionless modes.
 These result from the way we treat the moments in each sphere as independent degrees of freedom.
In the lower two modes, the moments in the two V and in their adjacent O stay locked to each other as in the reference configuration and precess together as expected for a single $S=1/2$ per rung  in a rigid spin approximation. 
In the higher modes, they precess independently of each other. In analogy to phonons, we label them as acoustic and optic modes respectively. For example, in the two modes near 250 meV, the two V across a bridge, precess opposite to each other. Inspection shows that there are four modes near 940 meV, with eigenvectors with primarily O$_v$ contribution and  two near 690 meV with primarily $O_b$ weight. Additional modes appear near 73 meV and 160 meV related to Na and O$_c$ if we include these even smaller magnetic moments. 
While the meaning of these optic spin wave modes thus becomes clear within the model, the applicability of the model itself  needs to be questioned. 
The distributed moment model used here describes a single $S=1/2$ electron in a molecular orbital centered on the bridge O$_b$ as a static probability distribution over the sites contributing to this orbital, but this represents in an approximate way the time-dependent fluctuations between the sites. 
The high energy modes become comparable in energy to  the longitudinal energy scale and the adiabatic decoupling underlying the validity of the Heisenberg Hamiltonian breaks down.
Estimating the local longitudinal energy scale by means of a constrained moment calculation, which quenches the small moments on O$_v$ and O$_b$ we find that this costs an energy
$\delta E=m\delta B+\frac{1}{2}\chi(\delta B)^2$ of order 20-30 meV per site. 
Here, $\delta B$ is the local  magnetic field along the spin direction required to quench the moment $m$ and $\chi$ is the longitudinal suscptibility on a specific site. This longitudinal energy scale is  small (instead of large as required by the adiabatic approximation) compared to the spin wave energies of the high energy modes  dominated by the small moment precession of the O$_v$ or O$_b$, indicating that we cannot  treat these as independent dynamical degrees of freedom. In fact, it costs less energy to quench these small induced moments on the O than to tilt them away from their antiferromagnentic alignment with V moments in a transverse mode.  
Similarly the 250 meV mode corresponding to the out of phase precession of the moments on the two V across a rung should be compared to the charge disproportionation mode which localizes the spin more on one V than the other. The total energy  difference between the $Pmmn$ and $P2_1mn$ structures per atom is of order 0.1 eV or less per atom and is thus also lower than this optic spin-wave excitation. Hence, we conclude that in a theory coupling longitudinal and charge excitations to the trasnverse excitations, these high optic spin wave excitations would be suppressed.

The lower two branches show a sine-like dispersion along $\Gamma Y$ with a maximum value of 53 meV at $Y$ for the second branch and 28 meV for the lower branch. These correspond to two weakly interacting chains of V-O$_b$-V rungs  in the cell with the usual $|2J_{eff}\sin{(q_yb/2)}|$ behavior of a single antiferromagnetic chain with $J_{eff}\approx-20$ meV, consistent with \cite{Isobe1996}. 
However,  the lowest mode becomes imaginary near ${\bf q}=0$ and stays imaginary along $\Gamma Z U$. This indicates an instability of the reference system towards such spin waves. The imaginary eigenvalue for $k_z\ne 0$ is caused by the mostly negative   exchange interactions between the layers while the reference spin configuration has parallel spins in adjacent layers. While the low energy state below 34 K is indeed known to involve several layers in terms of its charge disproportionation ordering, the spin ordering among layers is presently unknown but here predicted to involve spin alternation between adjacent layers. 

Inelastic neutron scattering measurements \cite{Grenier2001,Yoshihama1998} of the spin wave dispersion
have been studied only below the charge disproportionation critical temperature of 34 K and revealed two branches with slightly different spin gap
near 9.8 meV. The existence of a gap was attributed to the spin-Peierls dimerization along the chain, and the splitting of this gap along $\Gamma X$ was attributed to the charge disproportionation\cite{Valenti2001}. While our present calculation does not  correspond to the low
temperature charge disproportionated structure, and is classical in nature,  and hence does not show the spin gap addressed in \cite{Valenti2001}, we may still attempt to compare our spin wave dispersion for the acoustic spin waves to their data. The spin wave maximum at the Brillouin zone boundary in the {\bf b} direction was reported as 93$\pm$6 meV in \cite{Grenier2001} and 59 meV in \cite{Yoshihama1998} while our value is $\sim50$ meV. Given the uncertainty in the experimental extrapolation beyond the small wavevector range over which it was measured, this may be  considered reasonable agreement and again indicates that our distributed moment approach in an approximate way reproduces the overall effective energy scale of the acoustic spin waves.  

In conclusion, the distributed moment approach presented here provides an alternative point of  view for  a single spin in a molecular type orbital, 
which satisfactorily explains the scale of exchange interactions. It provides good agreement for the magnetic ordering temperature of about 320 K 
and the slope of the low energy spin waves or spin-stiffnes. Surprisingly, within this model the oxygen-vanadium exchange interactions play a crucial role in spite of the  smallness of the magnetic moments induced on the oxygens.

{\sl Methods:} The calculations were performed using the Full-Potential Linearized Muffin-Tin Orbital basis set as implemented in the {\sc Questaal}-suite of codes \cite{questaalpaper} which supports density functional theory and  QS$GW$ many-body perturbation theory as well as the linear response method of \cite{Kotani2008} for the transverse spin susceptibility. Details of the methodology and the spin wave and $T_c$ analytical calculations are given in \cite{Garcia2026arxivlinav2o5-spin}.

{\sl Acknowledgments:} This work was supported by the U.S. Department of Energy Basic Energy Sciences (DOE-BES) under Grant No. DE-SC0008933. Calculations made use of the High-Performance Computing Resource in the Core Facility for Advanced Research Computing at Case Western Reserve University, the Ohio Supercomputer Center, and the National Energy Research Scientific Computing Center, a DOE Office of Science user facility supported under Award No. DE-AC02-05CH11231 using NERSC Award No. BES- ERCAP0021783. C.G.'s stay at NREL (now NLR) was supported by the Science Graduate Student Research Program. S.A. and M.v.S. are supported by the Computational Chemical Sciences program within the U.S. DOE, Office of Science, BES, under Award No. DE-AC36-08GO28308. J.J. acknowledges support under the CCP9 project Computational Electronic Structure of Condensed Matter [part of the Computational Science Centre for Research Communities (CoSeC).  

  {\sl DataAvailability:} The digital data related to this paper are publicly available in \cite{DATANaV2O5}.

\bibliography{../Bib/linav2o5,../Bib/lmto,../Bib/gw,../Bib/dft,../Bib/spinchi}

\begin{thebibliography}{37}%
\makeatletter
\providecommand \@ifxundefined [1]{%
 \@ifx{#1\undefined}
}%
\providecommand \@ifnum [1]{%
 \ifnum #1\expandafter \@firstoftwo
 \else \expandafter \@secondoftwo
 \fi
}%
\providecommand \@ifx [1]{%
 \ifx #1\expandafter \@firstoftwo
 \else \expandafter \@secondoftwo
 \fi
}%
\providecommand \natexlab [1]{#1}%
\providecommand \enquote  [1]{``#1''}%
\providecommand \bibnamefont  [1]{#1}%
\providecommand \bibfnamefont [1]{#1}%
\providecommand \citenamefont [1]{#1}%
\providecommand \href@noop [0]{\@secondoftwo}%
\providecommand \href [0]{\begingroup \@sanitize@url \@href}%
\providecommand \@href[1]{\@@startlink{#1}\@@href}%
\providecommand \@@href[1]{\endgroup#1\@@endlink}%
\providecommand \@sanitize@url [0]{\catcode `\\12\catcode `\$12\catcode
  `\&12\catcode `\#12\catcode `\^12\catcode `\_12\catcode `\%12\relax}%
\providecommand \@@startlink[1]{}%
\providecommand \@@endlink[0]{}%
\providecommand \url  [0]{\begingroup\@sanitize@url \@url }%
\providecommand \@url [1]{\endgroup\@href {#1}{\urlprefix }}%
\providecommand \urlprefix  [0]{URL }%
\providecommand \Eprint [0]{\href }%
\providecommand \doibase [0]{https://doi.org/}%
\providecommand \selectlanguage [0]{\@gobble}%
\providecommand \bibinfo  [0]{\@secondoftwo}%
\providecommand \bibfield  [0]{\@secondoftwo}%
\providecommand \translation [1]{[#1]}%
\providecommand \BibitemOpen [0]{}%
\providecommand \bibitemStop [0]{}%
\providecommand \bibitemNoStop [0]{.\EOS\space}%
\providecommand \EOS [0]{\spacefactor3000\relax}%
\providecommand \BibitemShut  [1]{\csname bibitem#1\endcsname}%
\let\auto@bib@innerbib\@empty
\bibitem [{\citenamefont {Isobe}\ and\ \citenamefont {Ueda}(1996)}]{Isobe1996}%
  \BibitemOpen
  \bibfield  {author} {\bibinfo {author} {\bibfnamefont {M.}~\bibnamefont
  {Isobe}}\ and\ \bibinfo {author} {\bibfnamefont {Y.}~\bibnamefont {Ueda}},\
  }\bibfield  {title} {\bibinfo {title} {{Magnetic Susceptibility of
  Quasi-One-Dimensional Compound $\alpha'$-NaV2O5 --Possible Spin-Peierls
  Compound with High Critical Temperature of 34 K}},\ }\href
  {https://doi.org/10.1143/JPSJ.65.1178} {\bibfield  {journal} {\bibinfo
  {journal} {Journal of the Physical Society of Japan}\ }\textbf {\bibinfo
  {volume} {65}},\ \bibinfo {pages} {1178} (\bibinfo {year}
  {1996})}\BibitemShut {NoStop}%
\bibitem [{\citenamefont {Carpy}\ \emph {et~al.}(1972)\citenamefont {Carpy},
  \citenamefont {Casalot}, \citenamefont {Pouchard}, \citenamefont {Galy},\
  and\ \citenamefont {Hagenmuller}}]{Carpy72}%
  \BibitemOpen
  \bibfield  {author} {\bibinfo {author} {\bibfnamefont {A.}~\bibnamefont
  {Carpy}}, \bibinfo {author} {\bibfnamefont {A.}~\bibnamefont {Casalot}},
  \bibinfo {author} {\bibfnamefont {M.}~\bibnamefont {Pouchard}}, \bibinfo
  {author} {\bibfnamefont {J.}~\bibnamefont {Galy}},\ and\ \bibinfo {author}
  {\bibfnamefont {P.}~\bibnamefont {Hagenmuller}},\ }\bibfield  {title}
  {\bibinfo {title} {{Propri\'et\'es electriques et magn\'etiques des bronzes
  oxyfluor\'es de vanadium de formule $\alpha^\prime$-NaV$_2$O$_{5-x}$F$_x$ $(0
  \le x \le 1)$}},\ }\href
  {https://doi.org/https://doi.org/10.1016/0022-4596(72)90033-3} {\bibfield
  {journal} {\bibinfo  {journal} {Journal of Solid State Chemistry}\ }\textbf
  {\bibinfo {volume} {5}},\ \bibinfo {pages} {229} (\bibinfo {year}
  {1972})}\BibitemShut {NoStop}%
\bibitem [{\citenamefont {Bonner}\ and\ \citenamefont
  {Fisher}(1964)}]{Bonnerfisher64}%
  \BibitemOpen
  \bibfield  {author} {\bibinfo {author} {\bibfnamefont {J.~C.}\ \bibnamefont
  {Bonner}}\ and\ \bibinfo {author} {\bibfnamefont {M.~E.}\ \bibnamefont
  {Fisher}},\ }\bibfield  {title} {\bibinfo {title} {{Linear Magnetic Chains
  with Anisotropic Coupling}},\ }\href
  {https://doi.org/10.1103/PhysRev.135.A640} {\bibfield  {journal} {\bibinfo
  {journal} {Phys. Rev.}\ }\textbf {\bibinfo {volume} {135}},\ \bibinfo {pages}
  {A640} (\bibinfo {year} {1964})}\BibitemShut {NoStop}%
\bibitem [{\citenamefont {Pytte}(1974)}]{Pytte74}%
  \BibitemOpen
  \bibfield  {author} {\bibinfo {author} {\bibfnamefont {E.}~\bibnamefont
  {Pytte}},\ }\bibfield  {title} {\bibinfo {title} {{Peierls instability in
  Heisenberg chains}},\ }\href {https://doi.org/10.1103/PhysRevB.10.4637}
  {\bibfield  {journal} {\bibinfo  {journal} {Phys. Rev. B}\ }\textbf {\bibinfo
  {volume} {10}},\ \bibinfo {pages} {4637} (\bibinfo {year}
  {1974})}\BibitemShut {NoStop}%
\bibitem [{\citenamefont {Von~Schnering}\ \emph {et~al.}(1998)\citenamefont
  {Von~Schnering}, \citenamefont {Grin}, \citenamefont {Kaupp}, \citenamefont
  {Somer}, \citenamefont {Kremer}, \citenamefont {Jepsen}, \citenamefont
  {Chatterji},\ and\ \citenamefont {Weiden}}]{VonSchnering98}%
  \BibitemOpen
  \bibfield  {author} {\bibinfo {author} {\bibfnamefont {H.}~\bibnamefont
  {Von~Schnering}}, \bibinfo {author} {\bibfnamefont {Y.}~\bibnamefont {Grin}},
  \bibinfo {author} {\bibfnamefont {M.}~\bibnamefont {Kaupp}}, \bibinfo
  {author} {\bibfnamefont {M.}~\bibnamefont {Somer}}, \bibinfo {author}
  {\bibfnamefont {R.}~\bibnamefont {Kremer}}, \bibinfo {author} {\bibfnamefont
  {O.}~\bibnamefont {Jepsen}}, \bibinfo {author} {\bibfnamefont
  {T.}~\bibnamefont {Chatterji}},\ and\ \bibinfo {author} {\bibfnamefont
  {M.}~\bibnamefont {Weiden}},\ }\bibfield  {title} {\bibinfo {title}
  {{Redetermination of the crystal structure of sodium vanadate,
  $\alpha'$-NaV2O5}},\ }\href
  {https://www.scopus.com/inward/record.uri?eid=2-s2.0-0041138129&partnerID=40&md5=5d724ec0f44e14de4240725335222642}
  {\bibfield  {journal} {\bibinfo  {journal} {Zeitschrift fur Kristallographie
  - New Crystal Structures}\ }\textbf {\bibinfo {volume} {213}},\ \bibinfo
  {pages} {246} (\bibinfo {year} {1998})}\BibitemShut {NoStop}%
\bibitem [{\citenamefont {Smolinski}\ \emph {et~al.}(1998)\citenamefont
  {Smolinski}, \citenamefont {Gros}, \citenamefont {Weber}, \citenamefont
  {Peuchert}, \citenamefont {Roth}, \citenamefont {Weiden},\ and\ \citenamefont
  {Geibel}}]{Smolinski1998}%
  \BibitemOpen
  \bibfield  {author} {\bibinfo {author} {\bibfnamefont {H.}~\bibnamefont
  {Smolinski}}, \bibinfo {author} {\bibfnamefont {C.}~\bibnamefont {Gros}},
  \bibinfo {author} {\bibfnamefont {W.}~\bibnamefont {Weber}}, \bibinfo
  {author} {\bibfnamefont {U.}~\bibnamefont {Peuchert}}, \bibinfo {author}
  {\bibfnamefont {G.}~\bibnamefont {Roth}}, \bibinfo {author} {\bibfnamefont
  {M.}~\bibnamefont {Weiden}},\ and\ \bibinfo {author} {\bibfnamefont
  {C.}~\bibnamefont {Geibel}},\ }\bibfield  {title} {\bibinfo {title}
  {{${\mathrm{NaV}}_{2}{O}_{5}$ as a Quarter-Filled Ladder Compound}},\ }\href
  {https://doi.org/10.1103/PhysRevLett.80.5164} {\bibfield  {journal} {\bibinfo
   {journal} {Phys. Rev. Lett.}\ }\textbf {\bibinfo {volume} {80}},\ \bibinfo
  {pages} {5164} (\bibinfo {year} {1998})}\BibitemShut {NoStop}%
\bibitem [{\citenamefont {Horsch}\ and\ \citenamefont
  {Mack}(1998)}]{Horsch1998}%
  \BibitemOpen
  \bibfield  {author} {\bibinfo {author} {\bibfnamefont {P.}~\bibnamefont
  {Horsch}}\ and\ \bibinfo {author} {\bibfnamefont {F.}~\bibnamefont {Mack}},\
  }\bibfield  {title} {\bibinfo {title} {{A new view of the electronic
  structure of the spin-Peierls compound -NaVO}},\ }\href
  {https://doi.org/10.1007/s100510050455} {\bibfield  {journal} {\bibinfo
  {journal} {The European Physical Journal B - Condensed Matter and Complex
  Systems}\ }\textbf {\bibinfo {volume} {5}},\ \bibinfo {pages} {367} (\bibinfo
  {year} {1998})}\BibitemShut {NoStop}%
\bibitem [{\citenamefont {Storchak}\ \emph {et~al.}(2012)\citenamefont
  {Storchak}, \citenamefont {Parfenov}, \citenamefont {Eshchenko},
  \citenamefont {Lichti}, \citenamefont {Mengyan}, \citenamefont {Isobe},\ and\
  \citenamefont {Ueda}}]{Ueda2012}%
  \BibitemOpen
  \bibfield  {author} {\bibinfo {author} {\bibfnamefont {V.~G.}\ \bibnamefont
  {Storchak}}, \bibinfo {author} {\bibfnamefont {O.~E.}\ \bibnamefont
  {Parfenov}}, \bibinfo {author} {\bibfnamefont {D.~G.}\ \bibnamefont
  {Eshchenko}}, \bibinfo {author} {\bibfnamefont {R.~L.}\ \bibnamefont
  {Lichti}}, \bibinfo {author} {\bibfnamefont {P.~W.}\ \bibnamefont {Mengyan}},
  \bibinfo {author} {\bibfnamefont {M.}~\bibnamefont {Isobe}},\ and\ \bibinfo
  {author} {\bibfnamefont {Y.}~\bibnamefont {Ueda}},\ }\bibfield  {title}
  {\bibinfo {title} {{Antiferromagnetism in the spin-gap system
  ${\mathbf{NaV}}_{\mathbf{2}}{\mathbf{O}}_{\mathbf{5}}$: Muon spin rotation
  measurements}},\ }\href {https://doi.org/10.1103/PhysRevB.85.094406}
  {\bibfield  {journal} {\bibinfo  {journal} {Phys. Rev. B}\ }\textbf {\bibinfo
  {volume} {85}},\ \bibinfo {pages} {094406} (\bibinfo {year}
  {2012})}\BibitemShut {NoStop}%
\bibitem [{\citenamefont {Konstantinovi\ifmmode~\acute{c}\else \'{c}\fi{}}\
  \emph {et~al.}(2001)\citenamefont {Konstantinovi\ifmmode~\acute{c}\else
  \'{c}\fi{}}, \citenamefont {Dong}, \citenamefont {Ziaei}, \citenamefont
  {Clayman}, \citenamefont {Irwin}, \citenamefont {Yakushi}, \citenamefont
  {Isobe},\ and\ \citenamefont {Ueda}}]{Konstantinovic2001}%
  \BibitemOpen
  \bibfield  {author} {\bibinfo {author} {\bibfnamefont {M.~J.}\ \bibnamefont
  {Konstantinovi\ifmmode~\acute{c}\else \'{c}\fi{}}}, \bibinfo {author}
  {\bibfnamefont {J.}~\bibnamefont {Dong}}, \bibinfo {author} {\bibfnamefont
  {M.~E.}\ \bibnamefont {Ziaei}}, \bibinfo {author} {\bibfnamefont {B.~P.}\
  \bibnamefont {Clayman}}, \bibinfo {author} {\bibfnamefont {J.~C.}\
  \bibnamefont {Irwin}}, \bibinfo {author} {\bibfnamefont {K.}~\bibnamefont
  {Yakushi}}, \bibinfo {author} {\bibfnamefont {M.}~\bibnamefont {Isobe}},\
  and\ \bibinfo {author} {\bibfnamefont {Y.}~\bibnamefont {Ueda}},\ }\bibfield
  {title} {\bibinfo {title} {{Charge ordering and optical transitions of
  ${\mathrm{LiV}}_{2}{\mathrm{O}}_{5}$ and
  ${\mathrm{NaV}}_{2}{\mathrm{O}}_{5}$}},\ }\href
  {https://doi.org/10.1103/PhysRevB.63.121102} {\bibfield  {journal} {\bibinfo
  {journal} {Phys. Rev. B}\ }\textbf {\bibinfo {volume} {63}},\ \bibinfo
  {pages} {121102} (\bibinfo {year} {2001})}\BibitemShut {NoStop}%
\bibitem [{\citenamefont {Mostovoy}\ and\ \citenamefont
  {Khomskii}(1999)}]{Mostovoy1999}%
  \BibitemOpen
  \bibfield  {author} {\bibinfo {author} {\bibfnamefont {M.}~\bibnamefont
  {Mostovoy}}\ and\ \bibinfo {author} {\bibfnamefont {D.}~\bibnamefont
  {Khomskii}},\ }\bibfield  {title} {\bibinfo {title} {{Charge ordering and
  opening of spin gap in NaV2O5}},\ }\href
  {https://doi.org/https://doi.org/10.1016/S0038-1098(99)00453-6} {\bibfield
  {journal} {\bibinfo  {journal} {Solid State Communications}\ }\textbf
  {\bibinfo {volume} {113}},\ \bibinfo {pages} {159} (\bibinfo {year}
  {1999})}\BibitemShut {NoStop}%
\bibitem [{\citenamefont {Cuoco}\ \emph {et~al.}(1999)\citenamefont {Cuoco},
  \citenamefont {Horsch},\ and\ \citenamefont {Mack}}]{Cuoco1999}%
  \BibitemOpen
  \bibfield  {author} {\bibinfo {author} {\bibfnamefont {M.}~\bibnamefont
  {Cuoco}}, \bibinfo {author} {\bibfnamefont {P.}~\bibnamefont {Horsch}},\ and\
  \bibinfo {author} {\bibfnamefont {F.}~\bibnamefont {Mack}},\ }\bibfield
  {title} {\bibinfo {title} {{Theoretical study of the optical conductivity of
  ${\ensuremath{\alpha}}^{\ensuremath{'}}\ensuremath{-}{\mathrm{NaV}}_{2}{\mathrm{O}}_{5}$}},\
  }\href {https://doi.org/10.1103/PhysRevB.60.R8438} {\bibfield  {journal}
  {\bibinfo  {journal} {Phys. Rev. B}\ }\textbf {\bibinfo {volume} {60}},\
  \bibinfo {pages} {R8438} (\bibinfo {year} {1999})}\BibitemShut {NoStop}%
\bibitem [{\citenamefont {Presura}\ \emph {et~al.}(2000)\citenamefont
  {Presura}, \citenamefont {van~der Marel}, \citenamefont {Damascelli},\ and\
  \citenamefont {Kremer}}]{Presura2000}%
  \BibitemOpen
  \bibfield  {author} {\bibinfo {author} {\bibfnamefont {C.}~\bibnamefont
  {Presura}}, \bibinfo {author} {\bibfnamefont {D.}~\bibnamefont {van~der
  Marel}}, \bibinfo {author} {\bibfnamefont {A.}~\bibnamefont {Damascelli}},\
  and\ \bibinfo {author} {\bibfnamefont {R.~K.}\ \bibnamefont {Kremer}},\
  }\bibfield  {title} {\bibinfo {title} {{Low-temperature ellipsometry of
  ${\ensuremath{\alpha}}^{\ensuremath{'}}$-${\mathrm{NaV}}_{2}{\mathrm{O}}_{5}$}},\
  }\href {https://doi.org/10.1103/PhysRevB.61.15762} {\bibfield  {journal}
  {\bibinfo  {journal} {Phys. Rev. B}\ }\textbf {\bibinfo {volume} {61}},\
  \bibinfo {pages} {15762} (\bibinfo {year} {2000})}\BibitemShut {NoStop}%
\bibitem [{\citenamefont {Long}\ \emph {et~al.}(1999)\citenamefont {Long},
  \citenamefont {Zhu}, \citenamefont {Musfeldt}, \citenamefont {Wei},
  \citenamefont {Koo}, \citenamefont {Whangbo}, \citenamefont {Jegoudez},\ and\
  \citenamefont {Revcolevschi}}]{Long1999}%
  \BibitemOpen
  \bibfield  {author} {\bibinfo {author} {\bibfnamefont {V.~C.}\ \bibnamefont
  {Long}}, \bibinfo {author} {\bibfnamefont {Z.}~\bibnamefont {Zhu}}, \bibinfo
  {author} {\bibfnamefont {J.~L.}\ \bibnamefont {Musfeldt}}, \bibinfo {author}
  {\bibfnamefont {X.}~\bibnamefont {Wei}}, \bibinfo {author} {\bibfnamefont
  {H.-J.}\ \bibnamefont {Koo}}, \bibinfo {author} {\bibfnamefont {M.-H.}\
  \bibnamefont {Whangbo}}, \bibinfo {author} {\bibfnamefont {J.}~\bibnamefont
  {Jegoudez}},\ and\ \bibinfo {author} {\bibfnamefont {A.}~\bibnamefont
  {Revcolevschi}},\ }\bibfield  {title} {\bibinfo {title} {{Polarized optical
  reflectance and electronic band structure of
  ${\ensuremath{\alpha}}^{\ensuremath{'}}\ensuremath{-}{\mathrm{NaV}}_{2}{\mathrm{O}}_{5}$}},\
  }\href {https://doi.org/10.1103/PhysRevB.60.15721} {\bibfield  {journal}
  {\bibinfo  {journal} {Phys. Rev. B}\ }\textbf {\bibinfo {volume} {60}},\
  \bibinfo {pages} {15721} (\bibinfo {year} {1999})}\BibitemShut {NoStop}%
\bibitem [{\citenamefont {Fujii}\ \emph {et~al.}(1997)\citenamefont {Fujii},
  \citenamefont {Nakao}, \citenamefont {Yosihama}, \citenamefont {Nishi},
  \citenamefont {Nakajima}, \citenamefont {Kakurai}, \citenamefont {Isobe},
  \citenamefont {Ueda},\ and\ \citenamefont {Sawa}}]{Fujii97}%
  \BibitemOpen
  \bibfield  {author} {\bibinfo {author} {\bibfnamefont {Y.}~\bibnamefont
  {Fujii}}, \bibinfo {author} {\bibfnamefont {H.}~\bibnamefont {Nakao}},
  \bibinfo {author} {\bibfnamefont {T.}~\bibnamefont {Yosihama}}, \bibinfo
  {author} {\bibfnamefont {M.}~\bibnamefont {Nishi}}, \bibinfo {author}
  {\bibfnamefont {K.}~\bibnamefont {Nakajima}}, \bibinfo {author}
  {\bibfnamefont {K.}~\bibnamefont {Kakurai}}, \bibinfo {author} {\bibfnamefont
  {M.}~\bibnamefont {Isobe}}, \bibinfo {author} {\bibfnamefont
  {Y.}~\bibnamefont {Ueda}},\ and\ \bibinfo {author} {\bibfnamefont
  {H.}~\bibnamefont {Sawa}},\ }\bibfield  {title} {\bibinfo {title} {{New
  Inorganic Spin-Peierls Compound NaV 2O 5 Evidenced by X-Ray and Neutron
  Scattering}},\ }\href {https://doi.org/10.1143/JPSJ.66.326} {\bibfield
  {journal} {\bibinfo  {journal} {Journal of the Physical Society of Japan}\
  }\textbf {\bibinfo {volume} {66}},\ \bibinfo {pages} {326} (\bibinfo {year}
  {1997})}\BibitemShut {NoStop}%
\bibitem [{\citenamefont {L\"udecke}\ \emph {et~al.}(1999)\citenamefont
  {L\"udecke}, \citenamefont {Jobst}, \citenamefont {van Smaalen},
  \citenamefont {Morr\'e}, \citenamefont {Geibel},\ and\ \citenamefont
  {Krane}}]{Ludecke1999}%
  \BibitemOpen
  \bibfield  {author} {\bibinfo {author} {\bibfnamefont {J.}~\bibnamefont
  {L\"udecke}}, \bibinfo {author} {\bibfnamefont {A.}~\bibnamefont {Jobst}},
  \bibinfo {author} {\bibfnamefont {S.}~\bibnamefont {van Smaalen}}, \bibinfo
  {author} {\bibfnamefont {E.}~\bibnamefont {Morr\'e}}, \bibinfo {author}
  {\bibfnamefont {C.}~\bibnamefont {Geibel}},\ and\ \bibinfo {author}
  {\bibfnamefont {H.-G.}\ \bibnamefont {Krane}},\ }\bibfield  {title} {\bibinfo
  {title} {{Acentric Low-Temperature Superstructure of
  $\mathrm{NaV}{}_{2}O{}_{5}$}},\ }\href
  {https://doi.org/10.1103/PhysRevLett.82.3633} {\bibfield  {journal} {\bibinfo
   {journal} {Phys. Rev. Lett.}\ }\textbf {\bibinfo {volume} {82}},\ \bibinfo
  {pages} {3633} (\bibinfo {year} {1999})}\BibitemShut {NoStop}%
\bibitem [{\citenamefont {de~Boer}\ \emph {et~al.}(2000)\citenamefont
  {de~Boer}, \citenamefont {Meetsma}, \citenamefont {Baas},\ and\ \citenamefont
  {Palstra}}]{deBoer2000}%
  \BibitemOpen
  \bibfield  {author} {\bibinfo {author} {\bibfnamefont {J.~L.}\ \bibnamefont
  {de~Boer}}, \bibinfo {author} {\bibfnamefont {A.}~\bibnamefont {Meetsma}},
  \bibinfo {author} {\bibfnamefont {J.}~\bibnamefont {Baas}},\ and\ \bibinfo
  {author} {\bibfnamefont {T.~T.~M.}\ \bibnamefont {Palstra}},\ }\bibfield
  {title} {\bibinfo {title} {{Spin-Singlet Clusters in the Ladder Compound
  ${\mathrm{NaV}}_{2}{O}_{5}$}},\ }\href
  {https://doi.org/10.1103/PhysRevLett.84.3962} {\bibfield  {journal} {\bibinfo
   {journal} {Phys. Rev. Lett.}\ }\textbf {\bibinfo {volume} {84}},\ \bibinfo
  {pages} {3962} (\bibinfo {year} {2000})}\BibitemShut {NoStop}%
\bibitem [{\citenamefont {Ohama}\ \emph {et~al.}(1999)\citenamefont {Ohama},
  \citenamefont {Yasuoka}, \citenamefont {Isobe},\ and\ \citenamefont
  {Ueda}}]{Ohama1999}%
  \BibitemOpen
  \bibfield  {author} {\bibinfo {author} {\bibfnamefont {T.}~\bibnamefont
  {Ohama}}, \bibinfo {author} {\bibfnamefont {H.}~\bibnamefont {Yasuoka}},
  \bibinfo {author} {\bibfnamefont {M.}~\bibnamefont {Isobe}},\ and\ \bibinfo
  {author} {\bibfnamefont {Y.}~\bibnamefont {Ueda}},\ }\bibfield  {title}
  {\bibinfo {title} {{Mixed valency and charge ordering in
  ${\ensuremath{\alpha}}^{\ensuremath{'}}\ensuremath{-}{\mathrm{NaV}}_{2}{\mathrm{O}}_{5}$}},\
  }\href {https://doi.org/10.1103/PhysRevB.59.3299} {\bibfield  {journal}
  {\bibinfo  {journal} {Phys. Rev. B}\ }\textbf {\bibinfo {volume} {59}},\
  \bibinfo {pages} {3299} (\bibinfo {year} {1999})}\BibitemShut {NoStop}%
\bibitem [{\citenamefont {Nakao}\ \emph {et~al.}(2000)\citenamefont {Nakao},
  \citenamefont {Ohwada}, \citenamefont {Takesue}, \citenamefont {Fujii},
  \citenamefont {Isobe}, \citenamefont {Ueda}, \citenamefont {Zimmermann},
  \citenamefont {Hill}, \citenamefont {Gibbs}, \citenamefont {Woicik},
  \citenamefont {Koyama},\ and\ \citenamefont {Murakami}}]{Nakao2000}%
  \BibitemOpen
  \bibfield  {author} {\bibinfo {author} {\bibfnamefont {H.}~\bibnamefont
  {Nakao}}, \bibinfo {author} {\bibfnamefont {K.}~\bibnamefont {Ohwada}},
  \bibinfo {author} {\bibfnamefont {N.}~\bibnamefont {Takesue}}, \bibinfo
  {author} {\bibfnamefont {Y.}~\bibnamefont {Fujii}}, \bibinfo {author}
  {\bibfnamefont {M.}~\bibnamefont {Isobe}}, \bibinfo {author} {\bibfnamefont
  {Y.}~\bibnamefont {Ueda}}, \bibinfo {author} {\bibfnamefont {M.~v.}\
  \bibnamefont {Zimmermann}}, \bibinfo {author} {\bibfnamefont {J.~P.}\
  \bibnamefont {Hill}}, \bibinfo {author} {\bibfnamefont {D.}~\bibnamefont
  {Gibbs}}, \bibinfo {author} {\bibfnamefont {J.~C.}\ \bibnamefont {Woicik}},
  \bibinfo {author} {\bibfnamefont {I.}~\bibnamefont {Koyama}},\ and\ \bibinfo
  {author} {\bibfnamefont {Y.}~\bibnamefont {Murakami}},\ }\bibfield  {title}
  {\bibinfo {title} {{X-Ray Anomalous Scattering Study of a Charge-Ordered
  State in ${\mathrm{NaV}}_{2}{O}_{5}$}},\ }\href
  {https://doi.org/10.1103/PhysRevLett.85.4349} {\bibfield  {journal} {\bibinfo
   {journal} {Phys. Rev. Lett.}\ }\textbf {\bibinfo {volume} {85}},\ \bibinfo
  {pages} {4349} (\bibinfo {year} {2000})}\BibitemShut {NoStop}%
\bibitem [{\citenamefont {Mazurenko}\ \emph {et~al.}(2002)\citenamefont
  {Mazurenko}, \citenamefont {Lichtenstein}, \citenamefont {Katsnelson},
  \citenamefont {Dasgupta}, \citenamefont {Saha-Dasgupta},\ and\ \citenamefont
  {Anisimov}}]{Mazurenko2002}%
  \BibitemOpen
  \bibfield  {author} {\bibinfo {author} {\bibfnamefont {V.~V.}\ \bibnamefont
  {Mazurenko}}, \bibinfo {author} {\bibfnamefont {A.~I.}\ \bibnamefont
  {Lichtenstein}}, \bibinfo {author} {\bibfnamefont {M.~I.}\ \bibnamefont
  {Katsnelson}}, \bibinfo {author} {\bibfnamefont {I.}~\bibnamefont
  {Dasgupta}}, \bibinfo {author} {\bibfnamefont {T.}~\bibnamefont
  {Saha-Dasgupta}},\ and\ \bibinfo {author} {\bibfnamefont {V.~I.}\
  \bibnamefont {Anisimov}},\ }\bibfield  {title} {\bibinfo {title} {{Nature of
  insulating state in ${\mathrm{NaV}}_{2}{\mathrm{O}}_{5}$ above
  charge-ordering transition: A cluster dynamical mean-field study}},\ }\href
  {https://doi.org/10.1103/PhysRevB.66.081104} {\bibfield  {journal} {\bibinfo
  {journal} {Phys. Rev. B}\ }\textbf {\bibinfo {volume} {66}},\ \bibinfo
  {pages} {081104} (\bibinfo {year} {2002})}\BibitemShut {NoStop}%
\bibitem [{\citenamefont {Momma}\ and\ \citenamefont {Izumi}(2011)}]{VESTA}%
  \BibitemOpen
  \bibfield  {author} {\bibinfo {author} {\bibfnamefont {K.}~\bibnamefont
  {Momma}}\ and\ \bibinfo {author} {\bibfnamefont {F.}~\bibnamefont {Izumi}},\
  }\bibfield  {title} {\bibinfo {title} {{{\it VESTA3} for three-dimensional
  visualization of crystal, volumetric and morphology data}},\ }\href
  {https://doi.org/10.1107/S0021889811038970} {\bibfield  {journal} {\bibinfo
  {journal} {Journal of Applied Crystallography}\ }\textbf {\bibinfo {volume}
  {44}},\ \bibinfo {pages} {1272} (\bibinfo {year} {2011})}\BibitemShut
  {NoStop}%
\bibitem [{\citenamefont {Kotani}\ and\ \citenamefont {van
  Schilfgaarde}(2008)}]{Kotani2008}%
  \BibitemOpen
  \bibfield  {author} {\bibinfo {author} {\bibfnamefont {T.}~\bibnamefont
  {Kotani}}\ and\ \bibinfo {author} {\bibfnamefont {M.}~\bibnamefont {van
  Schilfgaarde}},\ }\bibfield  {title} {\bibinfo {title} {{Spin wave dispersion
  based on the quasiparticle self-consistent GW method: NiO, MnO and
  $\alpha$-MnAs}},\ }\href {https://doi.org/10.1088/0953-8984/20/29/295214}
  {\bibfield  {journal} {\bibinfo  {journal} {Journal of Physics: Condensed
  Matter}\ }\textbf {\bibinfo {volume} {20}},\ \bibinfo {pages} {295214}
  (\bibinfo {year} {2008})}\BibitemShut {NoStop}%
\bibitem [{\citenamefont {van Schilfgaarde}\ \emph {et~al.}(2006)\citenamefont
  {van Schilfgaarde}, \citenamefont {Kotani},\ and\ \citenamefont
  {Faleev}}]{MvSQSGWprl}%
  \BibitemOpen
  \bibfield  {author} {\bibinfo {author} {\bibfnamefont {M.}~\bibnamefont {van
  Schilfgaarde}}, \bibinfo {author} {\bibfnamefont {T.}~\bibnamefont
  {Kotani}},\ and\ \bibinfo {author} {\bibfnamefont {S.}~\bibnamefont
  {Faleev}},\ }\bibfield  {title} {\bibinfo {title} {{Quasiparticle
  Self-Consistent $GW$ Theory}},\ }\href
  {https://doi.org/10.1103/PhysRevLett.96.226402} {\bibfield  {journal}
  {\bibinfo  {journal} {Phys. Rev. Lett.}\ }\textbf {\bibinfo {volume} {96}},\
  \bibinfo {pages} {226402} (\bibinfo {year} {2006})}\BibitemShut {NoStop}%
\bibitem [{\citenamefont {Kotani}\ \emph {et~al.}(2007)\citenamefont {Kotani},
  \citenamefont {van Schilfgaarde},\ and\ \citenamefont {Faleev}}]{Kotani07}%
  \BibitemOpen
  \bibfield  {author} {\bibinfo {author} {\bibfnamefont {T.}~\bibnamefont
  {Kotani}}, \bibinfo {author} {\bibfnamefont {M.}~\bibnamefont {van
  Schilfgaarde}},\ and\ \bibinfo {author} {\bibfnamefont {S.~V.}\ \bibnamefont
  {Faleev}},\ }\bibfield  {title} {\bibinfo {title} {{Quasiparticle
  self-consistent {GW} method: A basis for the independent-particle
  approximation}},\ }\href {https://doi.org/10.1103/PhysRevB.76.165106}
  {\bibfield  {journal} {\bibinfo  {journal} {Phys.Rev. B}\ }\textbf {\bibinfo
  {volume} {76}},\ \bibinfo {eid} {165106} (\bibinfo {year}
  {2007})}\BibitemShut {NoStop}%
\bibitem [{\citenamefont {Hedin}(1965)}]{Hedin65}%
  \BibitemOpen
  \bibfield  {author} {\bibinfo {author} {\bibfnamefont {L.}~\bibnamefont
  {Hedin}},\ }\bibfield  {title} {\bibinfo {title} {New method for calculating
  the one-particle green's function with application to the electron-gas
  problem},\ }\href {https://doi.org/10.1103/PhysRev.139.A796} {\bibfield
  {journal} {\bibinfo  {journal} {Phys. Rev.}\ }\textbf {\bibinfo {volume}
  {139}},\ \bibinfo {pages} {A796} (\bibinfo {year} {1965})}\BibitemShut
  {NoStop}%
\bibitem [{\citenamefont {Hedin}\ and\ \citenamefont
  {Lundqvist}(1969)}]{Hedin69}%
  \BibitemOpen
  \bibfield  {author} {\bibinfo {author} {\bibfnamefont {L.}~\bibnamefont
  {Hedin}}\ and\ \bibinfo {author} {\bibfnamefont {S.}~\bibnamefont
  {Lundqvist}},\ }\bibfield  {title} {\bibinfo {title} {Effects of
  electron-electron and electron-phonon interactions on the one-electron states
  of solids},\ }in\ \href@noop {} {\emph {\bibinfo {booktitle} {Solid State
  Physics, Advanced in Research and Applications}}},\ Vol.~\bibinfo {volume}
  {23},\ \bibinfo {editor} {edited by\ \bibinfo {editor} {\bibfnamefont
  {F.}~\bibnamefont {Seitz}}, \bibinfo {editor} {\bibfnamefont
  {D.}~\bibnamefont {Turnbull}},\ and\ \bibinfo {editor} {\bibfnamefont
  {H.}~\bibnamefont {Ehrenreich}}}\ (\bibinfo  {publisher} {Academic Press},\
  \bibinfo {address} {New York},\ \bibinfo {year} {1969})\ pp.\ \bibinfo
  {pages} {1--181}\BibitemShut {NoStop}%
\bibitem [{\citenamefont {Antropov}\ \emph {et~al.}(1996)\citenamefont
  {Antropov}, \citenamefont {Katsnelson}, \citenamefont {Harmon}, \citenamefont
  {van Schilfgaarde},\ and\ \citenamefont {Kusnezov}}]{Antropov1996}%
  \BibitemOpen
  \bibfield  {author} {\bibinfo {author} {\bibfnamefont {V.~P.}\ \bibnamefont
  {Antropov}}, \bibinfo {author} {\bibfnamefont {M.~I.}\ \bibnamefont
  {Katsnelson}}, \bibinfo {author} {\bibfnamefont {B.~N.}\ \bibnamefont
  {Harmon}}, \bibinfo {author} {\bibfnamefont {M.}~\bibnamefont {van
  Schilfgaarde}},\ and\ \bibinfo {author} {\bibfnamefont {D.}~\bibnamefont
  {Kusnezov}},\ }\bibfield  {title} {\bibinfo {title} {{Spin dynamics in
  magnets: Equation of motion and finite temperature effects}},\ }\href
  {https://doi.org/10.1103/PhysRevB.54.1019} {\bibfield  {journal} {\bibinfo
  {journal} {Phys. Rev. B}\ }\textbf {\bibinfo {volume} {54}},\ \bibinfo
  {pages} {1019} (\bibinfo {year} {1996})}\BibitemShut {NoStop}%
\bibitem [{\citenamefont {Tyablikov}(1959)}]{Tyablikov1959}%
  \BibitemOpen
  \bibfield  {author} {\bibinfo {author} {\bibfnamefont {S.~V.}\ \bibnamefont
  {Tyablikov}},\ }\bibfield  {title} {\bibinfo {title} {{Retarded and advanced
  Green functions in the theory of ferromagnetism.}},\ }\href@noop {}
  {\bibfield  {journal} {\bibinfo  {journal} {Ukr. Mat. Zh.}\ }\textbf
  {\bibinfo {volume} {11}},\ \bibinfo {pages} {287} (\bibinfo {year}
  {1959})}\BibitemShut {NoStop}%
\bibitem [{\citenamefont {Callen}(1963)}]{Callen63}%
  \BibitemOpen
  \bibfield  {author} {\bibinfo {author} {\bibfnamefont {H.~B.}\ \bibnamefont
  {Callen}},\ }\bibfield  {title} {\bibinfo {title} {{Green Function Theory of
  Ferromagnetism}},\ }\href {https://doi.org/10.1103/PhysRev.130.890}
  {\bibfield  {journal} {\bibinfo  {journal} {Phys. Rev.}\ }\textbf {\bibinfo
  {volume} {130}},\ \bibinfo {pages} {890} (\bibinfo {year}
  {1963})}\BibitemShut {NoStop}%
\bibitem [{\citenamefont {Rusz}\ \emph {et~al.}(2005)\citenamefont {Rusz},
  \citenamefont {Turek},\ and\ \citenamefont {Divi\ifmmode~\check{s}\else
  \v{s}\fi{}}}]{Rusz2005}%
  \BibitemOpen
  \bibfield  {author} {\bibinfo {author} {\bibfnamefont {J.}~\bibnamefont
  {Rusz}}, \bibinfo {author} {\bibfnamefont {I.}~\bibnamefont {Turek}},\ and\
  \bibinfo {author} {\bibfnamefont {M.}~\bibnamefont
  {Divi\ifmmode~\check{s}\else \v{s}\fi{}}},\ }\bibfield  {title} {\bibinfo
  {title} {{Random-phase approximation for critical temperatures of collinear
  magnets with multiple sublattices: $\mathrm{Gd}X$ compounds
  $(X=\mathrm{Mg},\mathrm{Rh},\mathrm{Ni},\mathrm{Pd})$}},\ }\href
  {https://doi.org/10.1103/PhysRevB.71.174408} {\bibfield  {journal} {\bibinfo
  {journal} {Phys. Rev. B}\ }\textbf {\bibinfo {volume} {71}},\ \bibinfo
  {pages} {174408} (\bibinfo {year} {2005})}\BibitemShut {NoStop}%
\bibitem [{\citenamefont {Barker}\ and\ \citenamefont
  {Bauer}(2019)}]{Barker2019}%
  \BibitemOpen
  \bibfield  {author} {\bibinfo {author} {\bibfnamefont {J.}~\bibnamefont
  {Barker}}\ and\ \bibinfo {author} {\bibfnamefont {G.~E.~W.}\ \bibnamefont
  {Bauer}},\ }\bibfield  {title} {\bibinfo {title} {{Semiquantum thermodynamics
  of complex ferrimagnets}},\ }\href
  {https://doi.org/10.1103/PhysRevB.100.140401} {\bibfield  {journal} {\bibinfo
   {journal} {Phys. Rev. B}\ }\textbf {\bibinfo {volume} {100}},\ \bibinfo
  {pages} {140401(R)} (\bibinfo {year} {2019})}\BibitemShut {NoStop}%
\bibitem [{\citenamefont {Barker}\ \emph {et~al.}(2021)\citenamefont {Barker},
  \citenamefont {Pashov},\ and\ \citenamefont {Jackson}}]{Barker2021}%
  \BibitemOpen
  \bibfield  {author} {\bibinfo {author} {\bibfnamefont {J.}~\bibnamefont
  {Barker}}, \bibinfo {author} {\bibfnamefont {D.}~\bibnamefont {Pashov}},\
  and\ \bibinfo {author} {\bibfnamefont {J.}~\bibnamefont {Jackson}},\
  }\bibfield  {title} {\bibinfo {title} {{Electronic structure and finite
  temperature magnetism of yttrium iron garnet}},\ }\href
  {https://doi.org/10.1088/2516-1075/abd097} {\bibfield  {journal} {\bibinfo
  {journal} {Electronic Structure}\ }\textbf {\bibinfo {volume} {2}},\ \bibinfo
  {pages} {044002} (\bibinfo {year} {2021})}\BibitemShut {NoStop}%
\bibitem [{\citenamefont {Valent\'{\i}}\ \emph {et~al.}(2001)\citenamefont
  {Valent\'{\i}}, \citenamefont {Saha-Dasgupta}, \citenamefont {Alvarez},
  \citenamefont {Po\ifmmode \check{z}\else \v{z}\fi{}gaj\ifmmode \check{c}\else
  \v{c}\fi{}i\ifmmode~\acute{c}\else \'{c}\fi{}},\ and\ \citenamefont
  {Gros}}]{Valenti2001}%
  \BibitemOpen
  \bibfield  {author} {\bibinfo {author} {\bibfnamefont {R.}~\bibnamefont
  {Valent\'{\i}}}, \bibinfo {author} {\bibfnamefont {T.}~\bibnamefont
  {Saha-Dasgupta}}, \bibinfo {author} {\bibfnamefont {J.~V.}\ \bibnamefont
  {Alvarez}}, \bibinfo {author} {\bibfnamefont {K.}~\bibnamefont {Po\ifmmode
  \check{z}\else \v{z}\fi{}gaj\ifmmode \check{c}\else
  \v{c}\fi{}i\ifmmode~\acute{c}\else \'{c}\fi{}}},\ and\ \bibinfo {author}
  {\bibfnamefont {C.}~\bibnamefont {Gros}},\ }\bibfield  {title} {\bibinfo
  {title} {{Modeling the Electronic Behavior of $\gamma$-LiV$_{2}${O}$_{5}$: A
  Microscopic Study}},\ }\href {https://doi.org/10.1103/PhysRevLett.86.5381}
  {\bibfield  {journal} {\bibinfo  {journal} {Phys. Rev. Lett.}\ }\textbf
  {\bibinfo {volume} {86}},\ \bibinfo {pages} {5381} (\bibinfo {year}
  {2001})}\BibitemShut {NoStop}%
\bibitem [{\citenamefont {Grenier}\ \emph {et~al.}(2001)\citenamefont
  {Grenier}, \citenamefont {Cepas}, \citenamefont {Regnault}, \citenamefont
  {Lorenzo}, \citenamefont {Ziman}, \citenamefont {Boucher}, \citenamefont
  {Hiess}, \citenamefont {Chatterji}, \citenamefont {Jegoudez},\ and\
  \citenamefont {Revcolevschi}}]{Grenier2001}%
  \BibitemOpen
  \bibfield  {author} {\bibinfo {author} {\bibfnamefont {B.}~\bibnamefont
  {Grenier}}, \bibinfo {author} {\bibfnamefont {O.}~\bibnamefont {Cepas}},
  \bibinfo {author} {\bibfnamefont {L.~P.}\ \bibnamefont {Regnault}}, \bibinfo
  {author} {\bibfnamefont {J.~E.}\ \bibnamefont {Lorenzo}}, \bibinfo {author}
  {\bibfnamefont {T.}~\bibnamefont {Ziman}}, \bibinfo {author} {\bibfnamefont
  {J.~P.}\ \bibnamefont {Boucher}}, \bibinfo {author} {\bibfnamefont
  {A.}~\bibnamefont {Hiess}}, \bibinfo {author} {\bibfnamefont
  {T.}~\bibnamefont {Chatterji}}, \bibinfo {author} {\bibfnamefont
  {J.}~\bibnamefont {Jegoudez}},\ and\ \bibinfo {author} {\bibfnamefont
  {A.}~\bibnamefont {Revcolevschi}},\ }\bibfield  {title} {\bibinfo {title}
  {{Charge Ordering and Spin Dynamics in ${\mathrm{NaV}}_{2}{O}_{5}$}},\ }\href
  {https://doi.org/10.1103/PhysRevLett.86.5966} {\bibfield  {journal} {\bibinfo
   {journal} {Phys. Rev. Lett.}\ }\textbf {\bibinfo {volume} {86}},\ \bibinfo
  {pages} {5966} (\bibinfo {year} {2001})}\BibitemShut {NoStop}%
\bibitem [{\citenamefont {Yosihama}\ \emph {et~al.}(1998)\citenamefont
  {Yosihama}, \citenamefont {Nishi}, \citenamefont {Nakajima}, \citenamefont
  {Kakurai}, \citenamefont {Fujii}, \citenamefont {Isobe}, \citenamefont
  {Kagami},\ and\ \citenamefont {Ueda}}]{Yoshihama1998}%
  \BibitemOpen
  \bibfield  {author} {\bibinfo {author} {\bibfnamefont {T.}~\bibnamefont
  {Yosihama}}, \bibinfo {author} {\bibfnamefont {M.}~\bibnamefont {Nishi}},
  \bibinfo {author} {\bibfnamefont {K.}~\bibnamefont {Nakajima}}, \bibinfo
  {author} {\bibfnamefont {K.}~\bibnamefont {Kakurai}}, \bibinfo {author}
  {\bibfnamefont {Y.}~\bibnamefont {Fujii}}, \bibinfo {author} {\bibfnamefont
  {M.}~\bibnamefont {Isobe}}, \bibinfo {author} {\bibfnamefont
  {C.}~\bibnamefont {Kagami}},\ and\ \bibinfo {author} {\bibfnamefont
  {Y.}~\bibnamefont {Ueda}},\ }\bibfield  {title} {\bibinfo {title} {{Spin
  Dynamics in NaV$_2$O$_5$ -- Inelastic Neutron Scattering}},\ }\href
  {https://doi.org/10.1143/JPSJ.67.744} {\bibfield  {journal} {\bibinfo
  {journal} {Journal of the Physical Society of Japan}\ }\textbf {\bibinfo
  {volume} {67}},\ \bibinfo {pages} {744} (\bibinfo {year} {1998})}\BibitemShut
  {NoStop}%
\bibitem [{\citenamefont {Pashov}\ \emph {et~al.}(2019)\citenamefont {Pashov},
  \citenamefont {Acharya}, \citenamefont {Lambrecht}, \citenamefont {Jackson},
  \citenamefont {Belashchenko}, \citenamefont {Chantis}, \citenamefont
  {Jamet},\ and\ \citenamefont {van Schilfgaarde}}]{questaalpaper}%
  \BibitemOpen
  \bibfield  {author} {\bibinfo {author} {\bibfnamefont {D.}~\bibnamefont
  {Pashov}}, \bibinfo {author} {\bibfnamefont {S.}~\bibnamefont {Acharya}},
  \bibinfo {author} {\bibfnamefont {W.~R.}\ \bibnamefont {Lambrecht}}, \bibinfo
  {author} {\bibfnamefont {J.}~\bibnamefont {Jackson}}, \bibinfo {author}
  {\bibfnamefont {K.~D.}\ \bibnamefont {Belashchenko}}, \bibinfo {author}
  {\bibfnamefont {A.}~\bibnamefont {Chantis}}, \bibinfo {author} {\bibfnamefont
  {F.}~\bibnamefont {Jamet}},\ and\ \bibinfo {author} {\bibfnamefont
  {M.}~\bibnamefont {van Schilfgaarde}},\ }\bibfield  {title} {\bibinfo {title}
  {{Questaal: A package of electronic structure methods based on the linear
  muffin-tin orbital technique}},\ }\href
  {https://doi.org/https://doi.org/10.1016/j.cpc.2019.107065} {\bibfield
  {journal} {\bibinfo  {journal} {Computer Physics Communications}\ ,\ \bibinfo
  {pages} {107065}} (\bibinfo {year} {2019})}\BibitemShut {NoStop}%
\bibitem [{\citenamefont {Garcia}\ \emph {et~al.}(2026)\citenamefont {Garcia},
  \citenamefont {Turan}, \citenamefont {Jackson}, \citenamefont {Acharya},
  \citenamefont {van Schilfgaarde},\ and\ \citenamefont
  {Lambrecht}}]{Garcia2026arxivlinav2o5-spin}%
  \BibitemOpen
  \bibfield  {author} {\bibinfo {author} {\bibfnamefont {C.}~\bibnamefont
  {Garcia}}, \bibinfo {author} {\bibfnamefont {I.~K.}\ \bibnamefont {Turan}},
  \bibinfo {author} {\bibfnamefont {J.}~\bibnamefont {Jackson}}, \bibinfo
  {author} {\bibfnamefont {S.}~\bibnamefont {Acharya}}, \bibinfo {author}
  {\bibfnamefont {M.}~\bibnamefont {van Schilfgaarde}},\ and\ \bibinfo {author}
  {\bibfnamefont {W.~R.~L.}\ \bibnamefont {Lambrecht}},\ }\href
  {https://arxiv.org/abs/2610.24263} {\bibinfo {title} {{Charge
  disproportionation and spin ordering in LiV$_2$O$_5$ and NaV$_2$O$_5$: a
  distributed moment point of view}}} (\bibinfo {year} {2026}),\ \Eprint
  {https://arxiv.org/abs/2610.24263} {arXiv:2610.24263 [cond-mat.mtrl-sci]}
  \BibitemShut {NoStop}%
\bibitem [{DAT(2026)}]{DATANaV2O5}%
  \BibitemOpen
  \href {https://github.com/walterlambrecht/NaV2O5-distributed_moment} {}
  (\bibinfo {year} {2026}),\ \bibinfo {note} {{The digital data pertaining to
  this paper are available at
  \url{https://github.com/walterlambrecht/NaV2O5-distributed_moment}}}\BibitemShut
  {NoStop}%
\end{thebibliography}%
\end{document}